\documentclass[ms]{agujournal}

\begin{document}

%
%


\title{Unraveling the Mystery of Indian Summer Monsoon Prediction: Improved Estimate of Predictability Limit}

%
%

\authors{Subodh Kumar Saha\affil{1}, Anupam Hazra\affil{1}, Samir Pokhrel\affil{1}, Hemantkumar S. Chaudhari\affil{1}, K. Sujith\affil{1}, Archana Rai\affil{1}, Hasibur Rahaman\affil{2}\, and B. N. Goswami\affil{3}}

\affiliation{1}{Indian Institute of Tropical Meteorology, Pune, India.}
\affiliation{2}{Indian National Centre for Ocean Information Services, Hyderabad, India.}
\affiliation{3}{Cotton University, Guwahati, India.}


\correspondingauthor{Subodh Kumar Saha}{subodh@tropmet.res.in}


\begin{keypoints}
\item The observed link between synoptic variability and predictable modes suggest a high predictability of the ISMR.
\item CFSv2 with improved physics shows ISMR prediction skill higher than current estimate of potential predictability limit. 
\item Model shows $\sim$70\% of interannual variability of ISMR is predictable, which is much higher than earlier estimates ($\sim$45\%).
\end{keypoints}

%
%

\begin{abstract}

Large socio-economic impact of the Indian Summer Monsoon (ISM) extremes motivated numerous attempts at its long range prediction over the past century. However, a rather estimated low potential predictability limit (PPL) of seasonal prediction of the ISM, contributed significantly by \lq internal\rq\, interannual variability was considered insurmountable. Here we show that the \lq internal\rq\, variability contributed by the ISM sub-seasonal (synoptic + intra-seasonal) fluctuations, so far considered chaotic, is partly predictable as found to be tied to slowly varying forcing (e.g. El Ni\~no and Southern Oscillation). This provides a scientific basis for predictability of the ISM rainfall beyond the conventional estimates of PPL. We establish a much higher actual limit of predictability (r$\sim$0.82) through an extensive re-forecast experiment (1920 years of simulation) by improving two major physics in a global coupled climate model, which raises a hope for a very reliable dynamical seasonal ISM forecasting in the near future.


\end{abstract}

%
%

\section{Introduction}
\label{sec:intro}

The livelihood of about one fifth of the world's population living in South Asia thrives on regular arrival of the summer monsoon rainfall. The quantum of Indian summer monsoon rainfall (ISMR), however, varies from year-to-year, which has a significant impact on the country's economy, food production \citep{Gadgil06}, and availability of fresh water for drinking and industrial uses. Therefore, a reliable prediction of the ISMR one season in advance has tremendous value not only to the country's policy makers, but also to the farmers for planning crop management strategy. However, the seasonal prediction of the ISMR has remained one of the \lq grand challenge\rq\, problems in climate science over the past century \citep{Blanford1884, Gadgil05, Rajeevan12}, with the skill of the most current generation of climate models being sub-optimal \citep{Rajeevan12}. In the absence of the new knowledge that we are presenting here, various studies, including some of our own, have indicated that the ISMR inherently has a low predictability \citep{Webster98, Goswami98, Sperber13,  Goswami05, Goswami06c}. The low potential predictability of the ISMR estimated by previous studies \citep[e.g.][]{Kumar05, Rajeevan12} remained unquestioned because the skill of prediction of ISMR by all models, so far, remained below this limit. While the skill of seasonal prediction of the ISMR by climate models has improved from an older generation of models \citep{Kumar05} to a newer generation of models \citep{Rajeevan12}, it still remained \lq low\rq\, (correlation between observation and predictions being, r$\sim$0.4) and significantly below the conventional potential predictability limit (PPL, r$\sim$0.65). Here we report with the availability of new improved coupled ocean-atmosphere models that skill of large ensembles of retrospective forecasts of ISMR could be higher than the conventional estimates of PPL. This led us to question the sacrosanctness of the current estimate of PPL in the context of ISMR prediction. 


While the skill of prediction of the tropical climate, in general, is much higher than its extra-tropical counterpart \citep{Charney81}, the Indian summer monsoon (ISM) is considered a special tropical climate system, strongly influenced by sub-seasonal (intra-seasonal and synoptic) variability \citep{Webster98, Goswami05, Goswami06c} and that potentially limits its skill of seasonal prediction. A major component of the sub-seasonal variability is the Monsoon  Intra-seasonal Oscillations (MISOs; also known as active and break phases), which contribute significantly to the seasonal mean and often referred to as the building blocks of the ISM, is of a larger spatial scale (~8,000 km) \citep{Goswami01, Goswami06c}. On the other hand, the synoptic systems (lows, depressions and cyclonic storms) are of smaller spatial scale (~1000 km) and have a relatively smaller contribution to the  seasonal mean rainfall \citep{Sikka80a}. The contribution of the sub-seasonal variability to the seasonal mean is considered as \lq climate noise\rq\, and perceived to be unpredictable at long lead (e.g. a season) due to their origin from higher frequency chaotic component of the monsoon system, leading to the low potential predictability limit of the ISMR \citep{Palmer94, Webster98, Goswami06c}. Nevertheless, recent study shows that a fraction of sub-seasonal variability may be predictable as the statistics of the MISOs are also modulated by the ENSO \citep{Dwivedi15}.

Ocean-atmosphere interactions being key to both the inter-annual as well as the intra-seasonal variability of the ISM \citep{Webster98, Wang05, Sperber13}, a coupled atmosphere-ocean general circulation model (AOGCM) is a necessary tool for seasonal prediction of the ISMR. The simulation of regional ISM climate remained a major challenge with hardly any improvement of the persistent dry bias over the Indian landmass and wet bias over the Indian Ocean in the latest CMIP (Coupled Model Inter-comparison Project) models as compared to the earlier generation of models \citep{Sperber13}. Therefore, improvement of AOGCMs in simulating the south Asian monsoon precipitation climate is key to improvement of skill of seasonal prediction. While a fraction of the dry bias in CMIP5 models may arise from underestimation of variance of high frequency daily precipitation (likely to be related to underestimation of mesoscale and synoptic variances), most of the dry bias over land and wet bias over the ocean is likely to be related to the poor simulation of the northward propagation and intensity of the MISOs \citep{Goswami17}. Furthermore, the global teleconnections associated with slowly varying boundary conditions (i.e. sea surface temperature, snow, soil moisture) contributes significantly to the mean and variability of the ISMR and the models often have difficulties in simulating these relationship with reasonable skill.

The paradigm shift in our understanding of ISMR predictability evolved with our finding that the monsoon \lq climate noise\rq\, is not unpredictable as thought previously but a significant part of it is actually linked with the predictable modes (e.g. ENSO). We use extensive re-forecast experiment (equivalent of 1920 years of simulation) using four versions of a AOGCM and demonstrate that improvements in the simulation of sub-seasonal variability, particularly the synoptic variability increases the ISMR forecast skill. Here we provide a much awaited robust evidence in favor of the hypothesis that the ISM is a highly predictable system \citep{Charney81}. Section \ref{sec:method} describes the model, data and method used in this study. The main results are given in Section \ref{sec:result} and results are summarized in Section \ref{sec:summary}.

\section{Data and Methods}
\label{sec:method}

\subsection{Model and Re-forecast experiments}
Under the Monsoon Mission project (http://www.tropmet.res.in/monsoon/) of the Ministry of Earth Sciences, Government of India, the NCEP climate forecast system version 2 \citep[CFSv2;][]{SuSaha14} has been selected as a base model for the future improvement for a reliable ISMR prediction. While the standard version of the CFSv2 has a reasonable skill in seasonal prediction (r$\sim$0.55) \citep{Ramu16}, it is also known to have some significant systematic biases \citep{Saha13, Saha14, Hazra17}. A version of the model has been developed aiming to improve sub-seasonal simulations with an improved convection and microphysics parameterization \citep{Hazra17}, while another version has been developed aiming to improve the teleconnections with an improved snow scheme in the land surface model \citep{Saha17}. Further, improvements in snow and microphysics are combined in order to reap the benefits of both improvements. 

Similar to NCEP \citep[i.e.][]{SuSaha14}, model is initialized on every 15th, 20th, 25th February, 3rd March and four cycles in a day (i.e. 00, 06, 12 and 18 GMT) for the years 1981 to 2010 (30 years). Therefore, for each year, 16 member ensemble simulations are performed. Each ensemble member is integrated for a total of 9 months. In this study, initial conditions (1981-2010) are taken from NCEP Climate Forecast System Reanalysis \citep[CFSR;][]{SuSaha10}  (http://cfs.ncep.noaa.gov). Re-forecast experiments is carried out by employing the following four versions of the model:
 \begin{enumerate}
 \item{the original NCEP CFSv2 \citep[termed here as CONT;][]{SuSaha14}}
 \item{CFSv2 with improved snow physics \citep[termed here as SNOW;][]{Saha17}}
 \item{CFSv2 with improved microphysics \citep[termed here as MPHY;][]{Hazra17}}
 \item{CFSv2 with combined SNOW and MPHY (termed here as SNMP)}
 \end{enumerate}
Therefore, all together 4 models, 30 years of simulation and each with 16 ensemble member results into ($4\times 30\times 16)=1920$ years of simulation, which provide us the data for exploring the ISMR skill, as well as the limit of predictability.

\subsection{Estimates of the Limit of Predictability}
Two different methods are used for estimating potential predictability limit of the ISMR with the above mentioned model re-forecast data.

\textit{Perfect Model Correlation:} In the perfect model correlation method, the model is considered perfect and each ensemble member deviates from the others due to error in the initial conditions. ISMR from an ensemble member is correlated with the same from all remaining ensemble members and so on \citep[e.g.][]{Kumar05, Rajeevan12}.  

\textit{Analysis of Variance Method:} The potential predictability is also derived based on analysis of variance \citep[ANOVA;][]{Rowell95, Rowell98} technique. The ANOVA is one of the basic approaches to study the predictability of the seasonal mean with a large number of studies relying on this technique for estimating predictability \citep[e.g.][]{Kang06, Saha16b}.

\subsection{Low pass filter}
Fourier analysis is used for eliminating/filtering the desired frequency from the time series of the all India (66.5-100.5$^\circ$E, 6.5-32.5$^\circ$N only land region) and central India (74.5-86.5$^\circ$E, 16.5-26.5$^\circ$N only land region) averaged rainfall. The harmonics 0, 1, 2, 3 represent the mean, variations on 1, 1/2 and 1/3 year time scale respectively. While mean and first three harmonics together represent the annual cycle, the remaining harmonics together represent the sub-seasonal anomaly. Filtered anomaly is constructed in the following steps: \\
a) All India (AI) and Central India (CI) averaged daily observed rainfall from Indian Meteorological Department (IMD) is constructed.\\
b) The time series of daily rainfall anomaly are reconstructed back after removing the harmonics one by one in succession. Up to 150th harmonic (i.e. 2.4 days of periodicity) is removed successively. Thus, time series without 0-3rd, 0-4th, 0-5th, ....., 0-150th  harmonics represent time series of anomaly less than  91.2,  73.0, 60.8, ......., 2.4 periodicity respectively.

\subsection{Observed data}
The Hadley Centre Global Sea Ice and Sea Surface Temperature (HadISST, 1$^\circ$x1$^\circ$ ) daily global data for the period 1981-2010 are used here \citep{Rayner03}. Global monthly precipitation at 2.5$^\circ$ x 2.5$^\circ$ resolution by merging satellite and gauge observed rainfall \citep{Adler03} and daily gridded (1$^\circ$x1$^\circ$) IMD rainfall \citep{Rajeevan06} for the period 1981-2010 are used.

\section{Results}
\label{sec:result}
We investigate the contribution of the sub-seasonal variability to the interannual variability of the ISMR and its association (if any) with the predictable component of the natural modes of the variability, such as ENSO. In the tropics, cloud clusters are in general formed through small scale updrafts (few hundred kilometers), that are often organized through invisible planetary scale circulations. However, the invisible link between smaller and larger scale systems are not understood clearly. Nevertheless, the synoptic activities, which are of smaller scale and brings intense rainfall, are found to be associated with the planetary scale circulations like Madden-Julian Oscillation \citep{Liebmann94, Maloney00} and MISOs \citep{Goswami03}. In principle, the predictable modes, which are of planetary scale and evolve on longer time scale may leave their signature on the smaller scale events. As the ISMR has a strong teleconnections with the global predictable modes and the sub-seasonal variability are the building block of the monsoon, signature of these predictable modes may be also evident  in the synoptic and MISO events.

Considering the sub-seasonal fluctuations at three time bands: i) 3-7 days (synoptic), ii) 10-20 days (super-synoptic or higher MISO, and iii) 25-60 days (lower MISO), we use variance (i.e. vigor of fluctuations) of observed daily AI averaged rainfall in different bands as a metric of sub-seasonal activity. Low to high frequency components are removed one-by-one gradually and seasonal (JJAS) variance of filtered anomaly is correlated with ISMR anomaly. An important revelation coming out is that ISMR has the strongest correlation (a maximum of 0.63) with variance in the band of 3-5 day periods (Figure \ref{figone}). Since the maximum number of synoptic disturbances happens to be over the CI, correlation of CI averaged variance with ISMR is also consistent with that of AI. The fact that the vigor of synoptic disturbances and not only their frequency or distribution, that is strongly linked with  the year-to-year variability of the ISMR, is a new understanding. It is to be noted that ISMR is also moderately (weakly) correlated with MISO variance of 10-20 days (25-60 days) band. We also note that, the synoptic and 10-20 days bands of MISO explains about 25\% of the total variance individually. Despite the fact that the spatial structure of the synoptic systems and their contributions to the mean ISMR are relatively smaller (as compared to those of the MISO), the ISMR anomaly is primarily affected by the year-to-year variations of the synoptic activities.

Are the variances of one or more of these bands tied to any predictable climate signals like the ENSO? Correlations between variances of filtered rainfall anomalies in various bands and Ni\~no3 SST reveal a strong inverse relationship on synoptic scale (a minimum correlation of -0.51) as well as that of 10-20 days MISO band (a minimum correlation of -0.57; Figure \ref{figone}). Similarly, the filtered variance of CI rainfall has strongest correlation over 3-5 days band (a minimum of -0.58), suggesting a strong association between Pacific SST and  predominant synoptic rainfall over central India. The spatial structure of correlations between SST (2m air temperature) over ocean (land) and CI/AI variance at strongest correlation bands (at $<$ 5.2 days and $<$14.6 days band) indicates a large scale ENSO like teleconnection pattern (Figure \ref{figS1}). Therefore, the statistics of the sub-seasonal fluctuations, which is so far considered as climate noise, is actually associated with the predictable component of the global climate variability. This revelation provides scientific basis for a much higher predictability of the ISMR.

If ISM is a highly predictable system, then why all the dynamical forecast systems so far have been failed miserably or have shown limited success ? Large and persisting systematic biases in simulating the South Asian Monsoon by climate models \citep{Sperber13} made the current models not good enough for achieving the potential high skill. Also, it is a rather common problem in almost all global coupled models that over the ISM region, intense (light) rainfall events, primarily associated with the synoptic systems, are highly underestimated (overestimated) \citep{Goswami17}. However, enough attention has not been given to improve the rainfall distribution pattern, because synoptic and MISO are considered  not important as long as seasonal prediction of the ISMR is concerned. Therefore, a coupled Atmosphere-Ocean-General-Circulation-Model (AOGCM) with high fidelity in simulating the sub-seasonal variability, seasonal mean ISM and its global teleconnections is essential for a skillful prediction of the ISMR. We aimed at achieving this goal by improving the Climate Forecast System, version 2 (CFSv2) model \citep{SuSaha14} selected under the Monsoon Mission, based on extensive predictability studies and diagnosis of biases (described in section \ref{sec:method}). 


Conventional wisdom makes us believe that it is impossible for a dynamical prediction system to cross the PPL \citep{Rajeevan12, Kumar05}. Results from our model experiments are quite contrary to that. The estimate of the PPL based on the perfect model correlation method and that based on analysis of variance (ANOVA) method are rather close to each other and the actual skill of the ISMR (Figure \ref{figtwo}a) lies either at the upper edge (CONT=0.56;  SNOW=0.62) or exceeds the PPL (MPHY=0.71; SNMP=0.63). We note that, despite the models used here being imperfect, the prediction skill of the models exceeds the conventional PPL. This implies that the actual PPL could be much higher than 0.71. Therefore, neither ANOVA nor the perfect model correlation methods are able estimate the limit of predictability and the actual PPL likely to be much higher than the re-forecast skill in MPHY (i.e.  $>$ 0.71).

What then is the actual PPL? Here we propose a new measure of predictability, which is based on the actual forecast skill of a particular model. It is well known that forecast error grows due to imperfect physics and numerical method used as well as error in the initial conditions \citep[ICs;][]{Lorenz63, Lorenz69}. Further, among the ICs, initial errors also vary. Therefore, some of the ICs may lead to a very accurate/erroneous forecast. As a result, different forecaster with same model and ICs from the same source may end up with different skills just because of the choice of ICs and number of ensemble used for the forecast. In other words, a large number of initial conditions can generate a distribution of all possible forecast skills of a model. It may be noted that perfect initial conditions are never achievable. Therefore, the maximum skill achieved by a forecast system with a large set of initial conditions is likely to be always below the skill with that of perfect ICs. As these are actual correlation skill (correlation between observed and ensemble averaged ISMR), we define the maximum of these skills as the actual PPL.

In order to demonstrate it, all possible subsets of \lq n\rq\, ensemble members from a combination of 16 ensemble members are constructed (i.e. $^{16}C_n$ subsets, each with \lq n\rq\, ensemble member, where n=2 to 16). As an example, using 16 ensemble member only one combination (or subset) is possible ($^{16} C_{16}$). Similarly, using 15 ensemble members, 16 combination/subset is possible ($^{16} C_{15}$). Ensemble average rainfall of all possible combinations are correlated with the observed rainfall individually. Hence, all possible combinations of 2, 3, 4, 5, 6, 7, 8, 9, 10, 11, 12, 13, 14, 15, 16 ensembles (out of 16) will eventually generate 120, 560, 1820, 4368, 8008, 11440, 12870, 11440, 8008, 4368, 1820, 560, 120, 16, 1 subsets respectively and hence a similar number of correlation skills with the observed rainfall can be obtained.

The variations of minimum and maximum correlation skill as a function of ensemble members (Figure \ref{figtwo}b) lead to the following noteworthy conclusions: (i) the maximum actual ISMR skill or the actual PPL in the improved model (i.e. MPHY) can reach up to 0.82 with 3-8 ensemble average, (ii) systematic improvements of a model leads to a shift in both the maximum and minimum of potential skill to the higher side, (iii) the 16 ensemble member averaged re-forecast skill of each model lies well within maximum and minimum of the ALP. These results clearly demonstrate that right now about 70\% of inter-annual variability of the ISMR is predictable by the model, a much higher limit than the earlier estimate of limit of predictability  (i.e. about  45\%). We note that the increase in the minimum skill due to increase in the number of ensemble members is much faster than the change in the maximum skill with the size of the ensemble number. This implies that use of larger ensemble members may increase the confidence level in the forecast by ensuring a minimum level of skill with a given number of ensemble members. It is notable from the Figure \ref{figtwo}b that the maximum skill or the PPL is a stronger function of model improvements compared to its dependence on the ensemble size. However, whether this increase in the minimum skill with the size of ensemble members would continue beyond the size used here (i.e. 16) remains to be explored.

The observed relationship between ISMR/Ni\~no3 and the variance of filtered rainfall anomaly should be faithfully reproduced by the model in order to achieve such a high PPL.  On the synoptic time scale (3-7 days) the model is able to capture the relationship quite well, where MPHY outperforms all the models (Figures \ref{figthree}, \ref{figS1}). However, the performance of all models in the 10-20 and 25-60 days MISO bands are limited. It is to be noted that synoptic events have maximum contribution to the seasonal ISMR skill followed by MISOs of 10-20 and 25-60 days band. Improvement in contribution of MISOs to the seasonal anomaly will be key towards further improvements in ISMR skill in the model. The relationship between sub-seasonal variance and Ni\~no3 SST in models also corroborates the observation. At three month lead time, ISMR is predictable with a correlation skill of 0.71, which is in fact higher than prediction skill of Ni\~no3/Ni\~no3.4 (r$\sim$0.6; Table \ref{tab:cor}). This finding suggests that non-ENSO sources of predictability are also important contributors to the improved ISMR skill in the model. We find that seasonal forecast skill on smaller spatial scales (i.e. state or district level) also increase with the improvement of the skill of the ISMR, which are more valuable to the farmers for planning of crop and water management. The grid point correlation exceeding ~0.5 increases significantly with the ISMR skill (from 0.56 to 0.82; Figure \ref{figS3}).

\section{Discussion and Conclusions}
\label{sec:summary}
The tropics is known to be more predictable on seasonal time scale than its extra-tropical counterpart, with clear exception in case of the Indian summer monsoon. It has been believed that a significant part of the ISMR is unpredictable due to strong influence of the sub-seasonal variability, which are chaotic in nature. \citet{Goswami05} suggested that only about 50\% of interannual variability of the ISMR is predictable and the remaining part is \lq climate noise\rq\, (i.e. synoptic and MISO). Several past studies using state-of-the-art climate have estimated a rather low PPL (r$\sim$0.65) of the ISMR. Therefore, it was thought that it is not possible for a dynamical forecast system to cross this limit in terms of actual prediction skill. However, our model with improved physics has achieved ISMR correlation skill 0.71, which is above the PPL.

\begin{figure}[ht!]
\centering
\centerline{\includegraphics[height=3.5in]{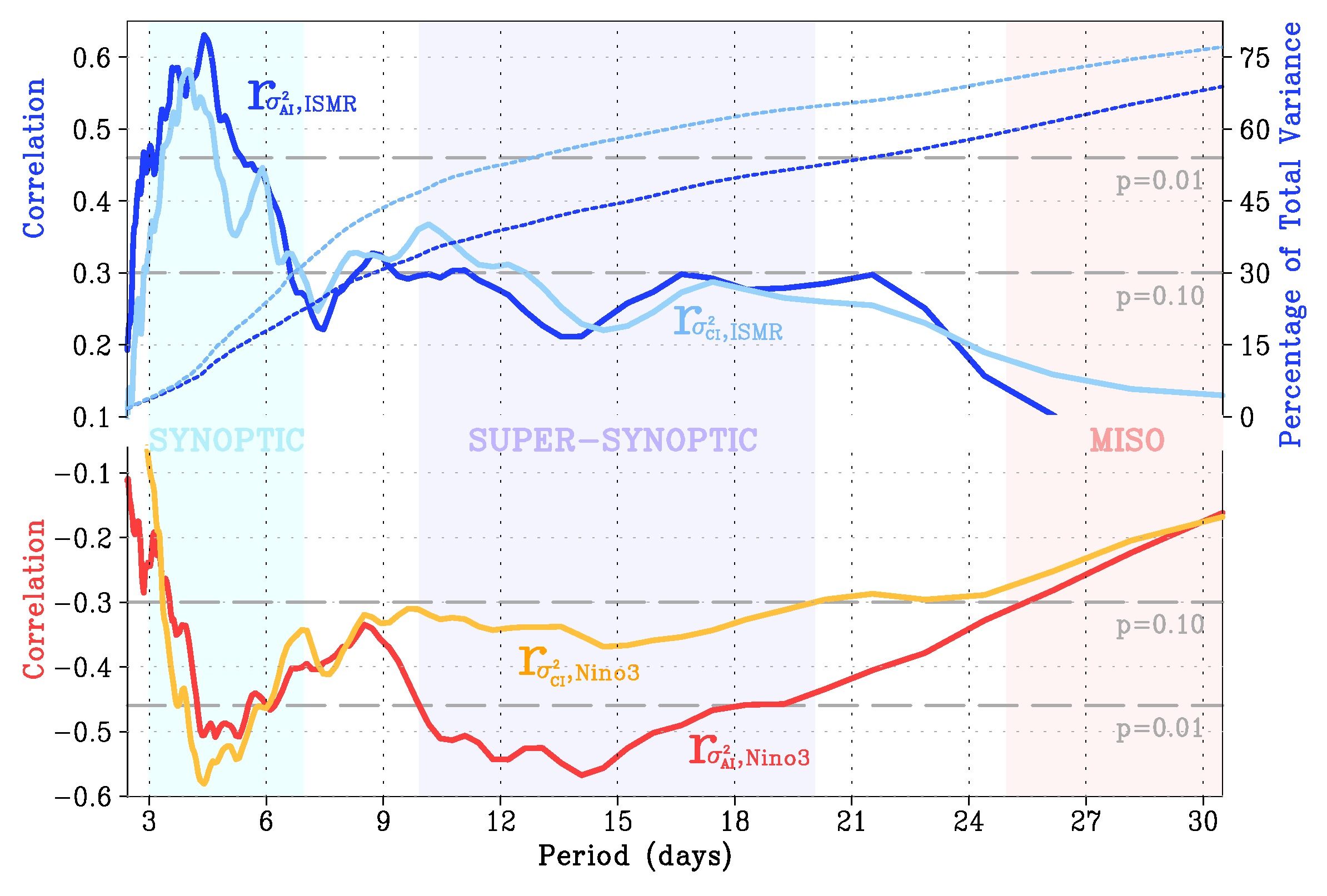}}
\caption{Seasonal (June to September average) ISMR and Ni\~no3 SST anomaly correlated with the sub-seasonal variance of rainfall at various time bands (or period). Correlation between variance of AI (CI) averaged filtered rainfall and ISMR anomaly is shown by solid deep blue (light blue) line.  Correlations between variance of AI (CI) rainfall and seasonal anomaly of Ni\~no3 SST are shown by red (orange) line. Dotted dark blue (light blue) shows percentage of filtered variance to the total variance of AI (CI) averaged rainfall.}
\label{figone}
\end{figure}

We found that the synoptic variability, which contributes about 25-30\% of the total sub-seasonal variance is predictable (Figure \ref{figone}). As the synoptic events are also known to be clustered with the active phases of the MISO \citep{Goswami03}, and a maximum predictability of the ISMR at less than 15 days is evident (Figure \ref{figone}), it is very likely that a part of high frequency MISO (10-20 days band and also known as super-synoptic) is also predictable. Therefore, in principle 70-80\% of the interannual variability of the ISMR is likely to be predictable. The newly found relationship is illustrated through schematic diagram (Figure \ref{figfive}). The Octopus is a \lq Big Brother\rq\, which consists of all natural predictable modes and controls synoptic and a part of the MISO on his own way. By keeping a tight lease on the statistics (e.g. variances) of the sub-seasonal fluctuations and their contribution to the seasonal mean, the ENSO effectively enhances the ISMR predictability significantly.

As the contribution of the sub-seasonal variance to the ISMR improves, models show improvements in the ISMR re-forecast skill (Figure \ref{figthree}). While observation shows a very weak relationship between ISMR and variance of lower MISOs (30-60 days), it is very strong positive in some models (CONT, SNOW). As a result, reasonable contribution by the synoptic events is nullified by unrealistic contributions of MISOs. This is also reflected in the actual skill (Figure \ref{figtwo}). Almost all global climate models show very poor performance in simulating the distribution of heavy and moderate to low rainfall events over the ISM region \citep{Goswami17}.  Furthermore, the synoptic system, which is primarily responsible for heavy rainfall event, is likely to change under the future global warming scenario \citep{Sandeep18}. Therefore, fidelity of a model to simulate the relationship between ISMR and sub-seasonal variability will define the reliability of the future scenario of the mean as well as predictability of the ISM.

In many previous studies it is found that often the skill of a model is greater than the PPL and that showed limitations of the methods for estimating PPL \citep[e.g.][]{Kumar14, Saha16}. Here we propose a new method for calculating the actual limit of predictability of the ISMR, which has a binding relationship with the actual observation. As the model improves (i.e. ISMR skill increases), the distribution pattern of actual correlation skill also moves to the higher side (Figure \ref{figtwo}) and the PPL scales up to 0.82 . Therefore, right now the model is able to predict about 70\% of the IAV, which is much higher than earlier estimates (i.e. about 45\%).

Further improvements in the contribution of sub-seasonal anomaly to the ISMR anomaly in the model may raise the PPL to around 0.9 ($\sim$80\% of observed variance), a target seems within reach in the near future. It appears that, further developments in model physics and ICs will take the forecast to such a level that at least sign of the ISMR anomaly (dry or wet year) is predicted without failure and amplitude with much greater confidence.

\acknowledgments
This work results from extensive model development activities at IITM under the Monsoon Mission project of the MoES, Govt. Of India. We thank MoES and Director IITM, HPCS for all the support to carry out this work. We also thank NCEP for providing initial conditions and modeling support. BNG is grateful to the Science and Engineering Research Board (SERB), Govt. India for a Fellowship. Authors duly acknowledge Mrs. Yashashri Rohan Jadav, Mantri Avenue-1, Panchavati, Pune for preparing schematic (Figure \ref{figfive}). The observational data sets used in the study are Hadley Centre Sea Ice and Sea Surface Temperature dataset \citep[HadISST, $1^\circ \times 1^\circ$;][]{Rayner03}. Gridded rainfall data taken from India Meteorological Department \citep[IMD; with $1^\circ \times 1^\circ$ horizontal resolution][]{Rajeevan06} as well as the Global Precipitation Climatology Project version 2 \citep[GPCP; $2.5^\circ \times 2.5^\circ$][]{Adler03}. Model re-forecast data is archived at IITM and can be accessed from the corresponding author upon request.

%
%
%
%
%
%
%
%
%


\pagebreak

\begin{table}
\caption{Correlations between observed (HadISST) and model SST over Ni\~no3 ($150^\circ$-$90^\circ$W, $5^\circ$S-$5^\circ$N) and Ni\~no3.4 ($170^\circ$-$120^\circ$W, $5^\circ$S-$5^\circ$N) regions and ensemble mean ISMR predicted by the four models.}
\centering
\begin{tabular}{c c c c}
\hline
  & Ni\~no3  &  Ni\~no3.4  \\ 
\hline
CONT    & 0.53  &  0.59 \\
SNOW    & 0.52  &  0.57 \\
MPHY    & 0.57  &  0.61 \\
SNMP    & 0.52  &  0.58 \\
\hline
\end{tabular}
\label{tab:cor}
\end{table}


\begin{figure}[ht!]
\centering
\centerline{\includegraphics[height=5.0in]{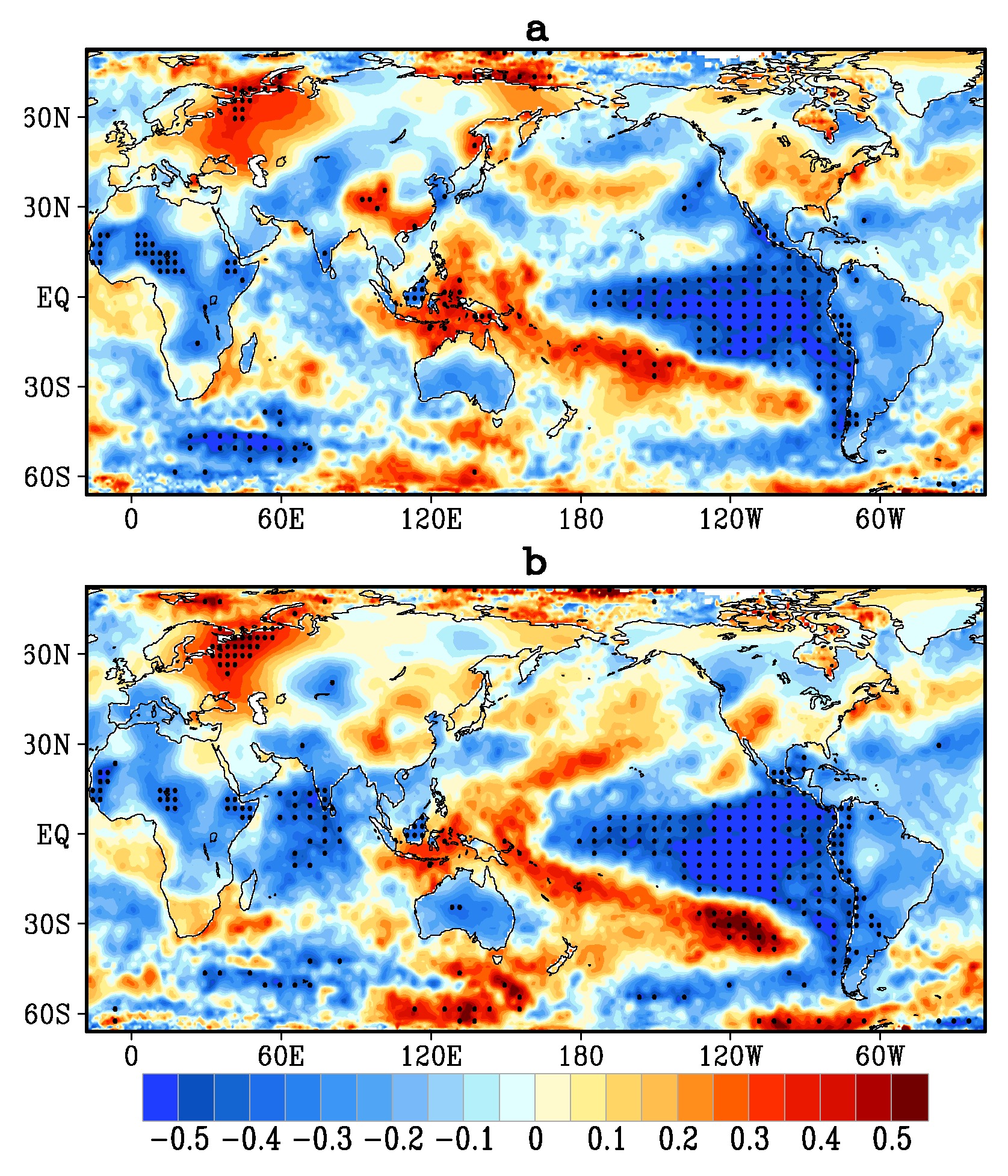}}
\caption{The AI averaged sub-seasonal rainfall variance below a particular periodicity is correlated with SST (2m temperature) at every grid point of global ocean (land).  a, Using seasonal variance of central India (CI) averaged rainfall $<$ 5.2 days periodicity. b, using seasonal variance of all India (AI) averaged rainfall $<$ 14.6 days periodicity. Correlations significant at 95\% level are stippled. The similarity of the patterns of correlation with the canonical ENSO SST confirms the modulation of the of sub-seasonal fluctuations over the ISM region through ENSO teleconnection.}
\label{figS1}
\end{figure}

\begin{figure}
\centering
\centerline{\includegraphics[height=3.0in]{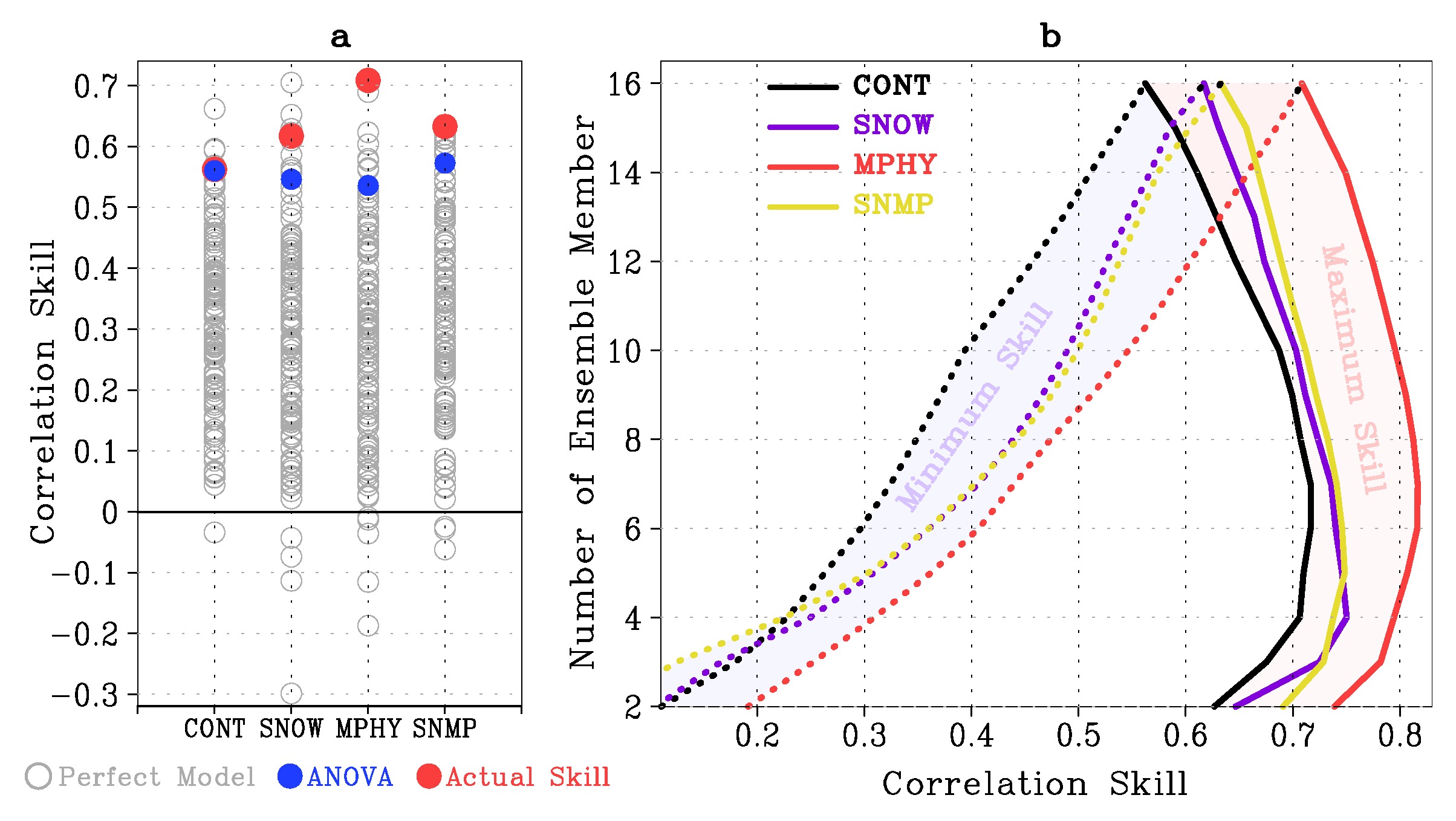}}
\caption{Limits of ISMR predictability and actual re-forecast skill. a, Potential skill of  the ISMR based on perfect model correlation (gray open circle),  ANOVA method (blue closed circle) and actual correlation skill between 16 ensemble averaged mean and Global Precipitation Climatology Project (GPCP) observation (red closed circle). b, The maximum (solid line) and minimum (dotted line) of actual correlation skill using all combination of \lq n\rq\, ensemble averaged ISMR (i.e. $^{16}C_n$ , where \lq n\rq\, varies from 2 to 16) from CONT (black), SNOW (purple), MPHY (red) and SNMP (yellow).}
\label{figtwo}
\end{figure}

\begin{figure}
\centering
\centerline{\includegraphics[height=3.5in]{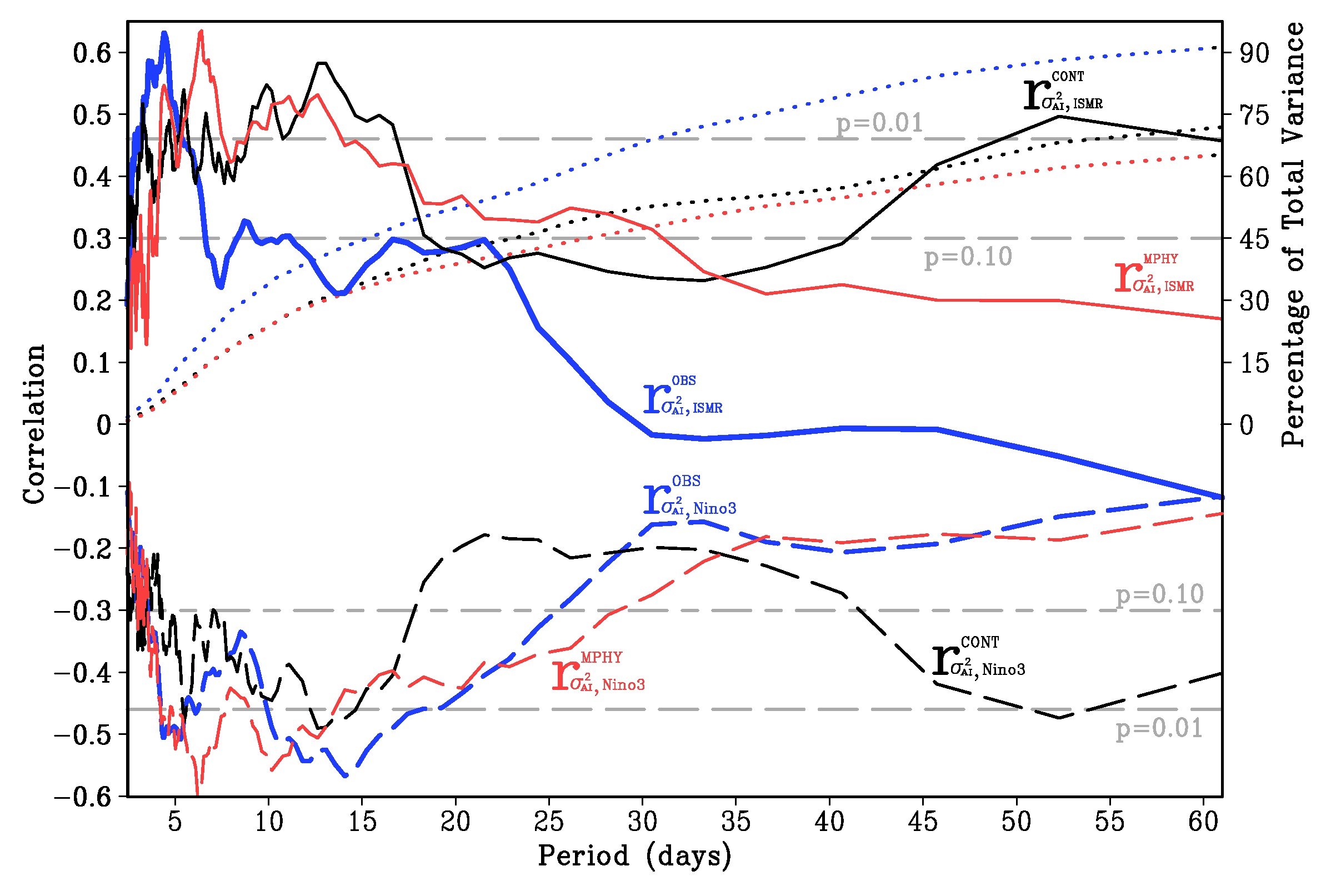}}
\caption{Same as Figure 1, but also using two model simulations. The relationship between filtered variances and seasonal anomaly of AI rain/Ni\~no3 SST from model having minimum (i.e. CONT, with black line) and maximum (i.e. MPHY, with red line) ISMR skill is presented along with the same from observation (blue line).}
\label{figthree}
\end{figure}

\begin{figure}
\centering
\centerline{\includegraphics[height=3.5in]{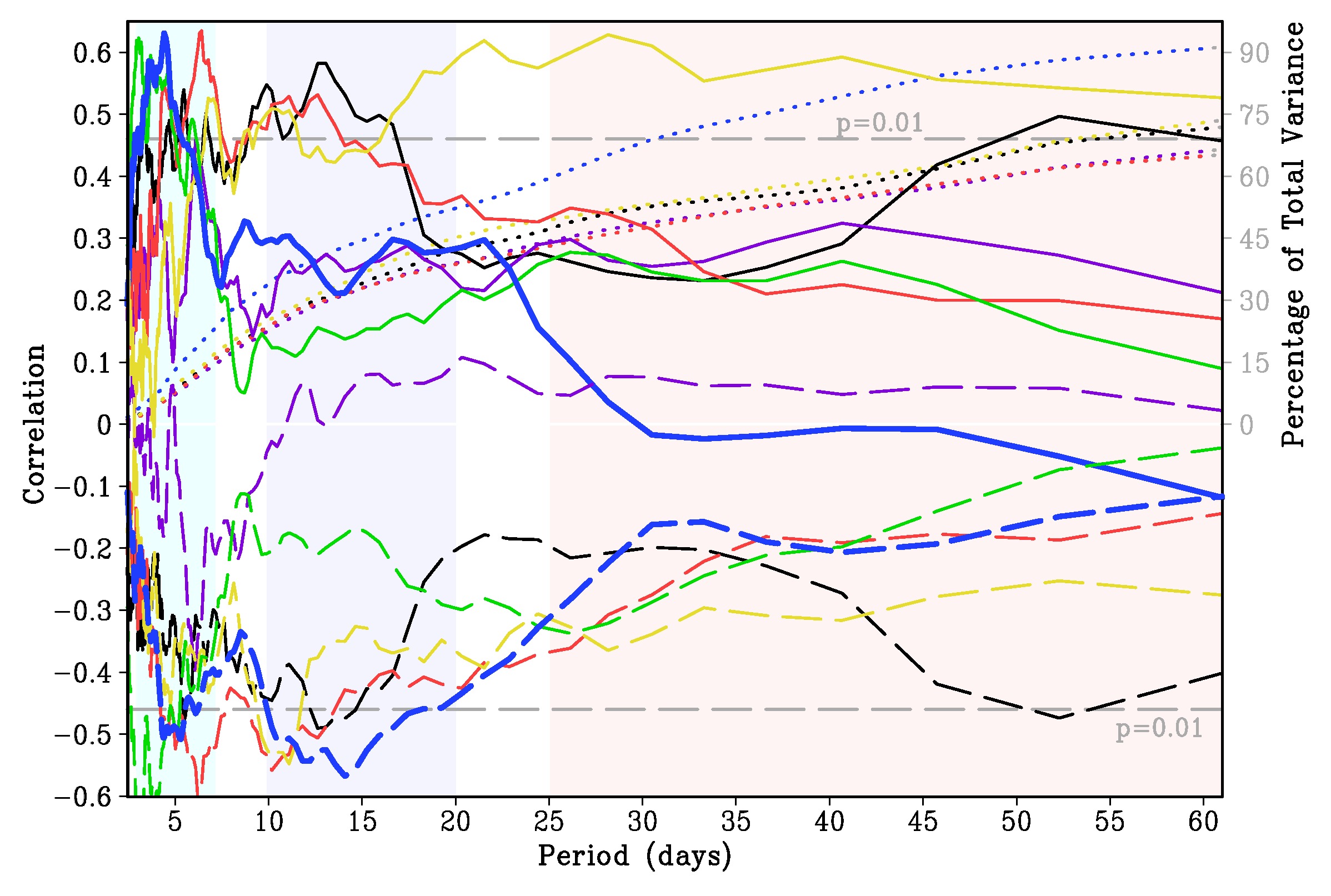}}
\caption{Same as Figure 1, but using observation and all four models. Blue, black, red, purple, and yellow lines correspond to observation (IMD, HadISST), CONT, MPHY, SNOW and SNMP respectively. The green line is for MPHY, with best 8 ensemble member resulting into maximum ISMR re-forecast skill (correlation = 0.82). The dotted lines represent percentage of total variance at various bands. Grey dashed lines represents correlation significant at 99\% level using two-tailed t-test.}
\label{figS1}
\end{figure}

\begin{figure}
\centering
\centerline{\includegraphics[height=2.5in]{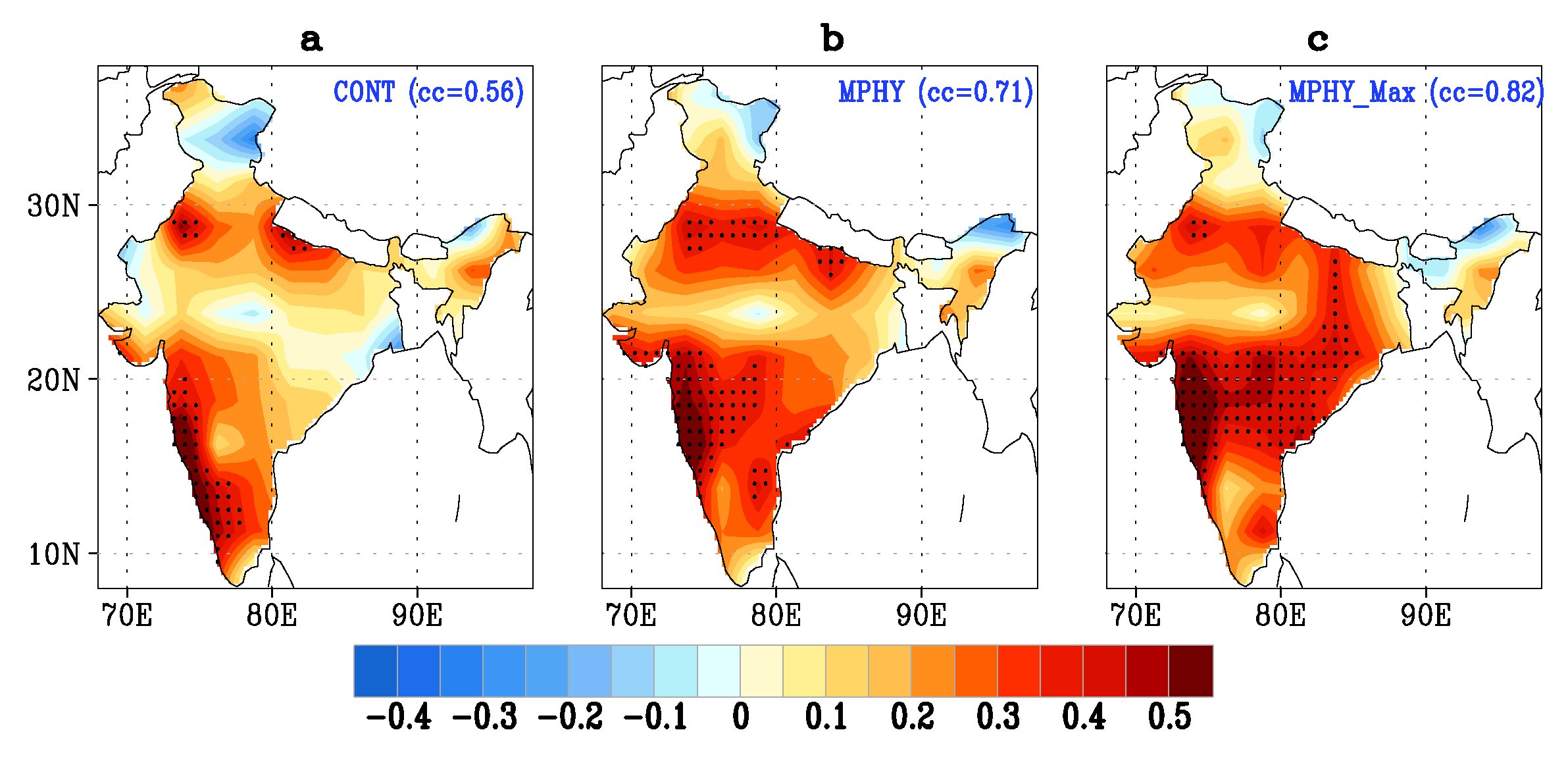}}
\caption{Grid point correlations between JJAS averaged GPCP rainfall and ensemble mean re-forecast. Correlation skill in (a) CONT, (b) MPHY and (c)  MPHY with maximum ALP using 8 ensemble member (MPHY\_Max). Correlation significant at 95\% (tow-tailed test) are stippled .}
\label{figS3}
\end{figure}

\begin{figure}
\centering
\centerline{\includegraphics[height=1.8in]{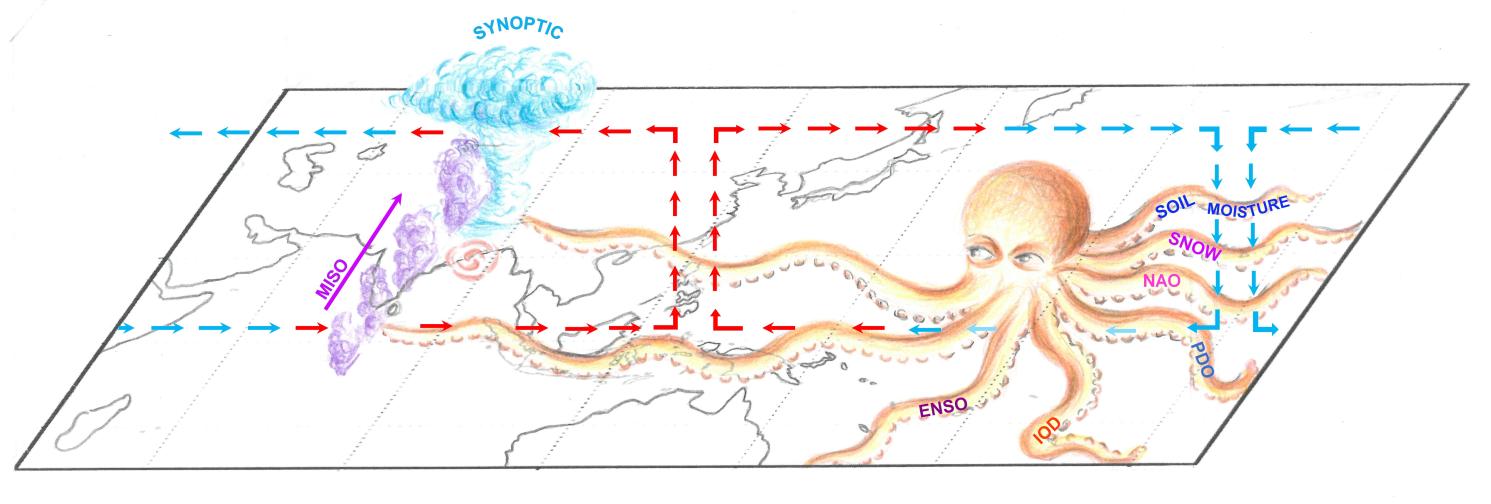}}
\caption{Schematic diagram illustrating the newly discovered association of large scale predictor (represented by a giant Octopus) with sub-seasonal component (synoptic + MISO) of the Indian summer monsoon. The predictor consists of all natural modes of variability, i.e. ENSO, Indian Ocean Dipole (IOD), Pacific Decadal Oscillation (PDO), North Atlantic Oscillation ( NAO), snow, soil moisture etc., each represents an arm of the predictor.  The two long arms of the predictor aiming towards the synoptic and MISO indicate its influences on these sub-seasonal processes. The large scale Walker circulation is shown by colored (red, blue) arrows.}
\label{figfive}
\end{figure}




\listofchanges

\end{document}